# Optimization and analysis of experimental parameters for polarization gradient cooling in optical molasses

Zhonghua Ji, Jinpeng Yuan, Yanting Zhao[a], Xuefang Chang, Yonggang Yang, Liantuan Xiao and Suotang Jia

*State Key Laboratory of Quantum Optics and Quantum Optics Devices, Laser Spectroscopy Laboratory, Shanxi University, Taiyuan 030006, China*

**Abstract**

We systematically investigate the dependence of the temperature of cold cesium atoms of polarization gradient cooling (PGC) in optical molasses on experimental parameters, which contain changing modes of cooling laser, PGC interaction time, cooling laser frequency and its intensity. The ′SR′ mode of cooling laser, that means the cooling laser frequency is changed with step mode and cooling laser intensity is changed with ramp mode, is found to be the best for PGC comparing with other ′SS′, ′RS′, and ′RR′ modes. We introduce a statistical explanation and an exponential decay function to explain the variation of cold atomic temperature on time. The heating effect is observed when the cooling laser intensity is lower than the saturation intensity of cold atoms. After optimization, the lowest temperature of cold cesium atoms is observed to be about 4μK with the number of $2\times10^9$, the density of $1\times10^{11}/cm^3$ and the phase space density of $4.4\times10^{-5}$. The optimization process and analysis of controllable experimental parameters are also meaningful for other cold atomic systems.

**PACS:** 37.10.De -Atom cooling methods, 37.10.Gh -Atom traps and guides, 67.85.-d – Ultracold gases, trapped gases

## 1. Introduction

With the development of laser cooling and trapping of neutral atoms, ultracold atoms have received increasing attentions due to their significant advances and potential applications, such as time and frequency standard[1], atom interferometry[2], Bose–Einstein condensation[3], atom laser[4], ultracold molecules[5], quantum information[6], and so on.

In order to achieve ultracold atoms, a variety of methods have been developed. Magneto-optical trap (MOT), as a simple and effective method, is widely used to cool and trap cold atoms from dilute atomic vapor at room temperature. However, the cold atomic temperature in MOT is around Doppler limit. In order to exceed the Doppler limit, sub-Doppler cooling is represented to achieve trapped atoms with much lower temperature. Polariztion gradient cooling (PGC) in optical molasses[7] is a widespread technique for its speed, simplify and effectiveness. The sub-Doppler cooling mechanism of cold atoms by PGC was very early proposed by Cohen-Tannoudji group[8] and Chu group[9]. Thence, researchers expanded this method to gray optical molasses[10,11], optical lattices[12], and even gray optical lattice[13]. Recent experiments represented that the PGC can also work well even for the small hyperfine splitting atoms[14,15].

---

[a] Email: zhaoyt@sxu.edu.cn

According to the theory of polarization gradient cooling, the cold atomic equilibrium temperature $T_e$ is given by

$$T_e \approx C\hbar\Omega^2 / k_B |\delta|, \qquad (1)$$

where $C$ is a number factor on the order of 0.1, $k_B$ is the Boltzmann constant, $\hbar$ is the Planck constant, $\delta=\omega_L-\omega_A$ is the laser detuning of the incident laser frequency $\omega_L$ from the atomic resonance frequency $\omega_A$, $\Omega=2\boldsymbol{d}\cdot\boldsymbol{E}/\hbar$ is the Rabi frequency describing the coupling between the atomic dipole moment $\boldsymbol{d}$ and the electric field vector $\boldsymbol{E}$. This equation means that the cold atomic temperature is inversely proportional to the cooling laser detuning at a given intensity and proportional to the cooling laser intensity at a given frequency. Many researchers focused on the dependence of cold atom temperature on the values of cooling laser intensity and detuning in final state. For examples, Salomon *et al.* found that the temperature of cold cesium atoms depended on the cooling laser intensity-to-detuning ratio[9]. Wang *et al.* measured the dependence of cold $^{87}$Rb atoms on the cooling laser intensity and detuning[16]. However, there are few literatures systematically investigating the influence of process of PGC on the cold atomic temperature[14,16].

In this paper, we will investigate the dependence of the temperature of cold cesium atoms on some important influencing experimental parameters, which contain the changing modes of cooling laser in PGC, the PGC interaction time, cooling laser frequency and its intensity. The process of analysis and optimization will help us to more deeply understand the interaction mechanism of cooling laser and cold atoms in PGC.

## 2. Experimental setup and timing sequence

Our experiment starts from a standard vapor-loaded $^{133}$Cs MOT in a commercially available stainless steel chamber[17]. The vacuum background pressure is usually about 5×10$^{-7}$Pa when atomic vapor fills the vacuum chamber. A pair of coils with anti-Helmholtz configuration generates about 12G/cm magnetic gradient, and three other pairs of coils with Helmholtz configuration are placed around the vacuum chamber to compensate geomagnetism at the position of cold atom clouds. Our laser setup consists of two 852 nm diode lasers, which are locked by saturated absorption spectroscopy (SAS). The first laser frequency is locked to $F=4\rightarrow F'=4$ transition by SAS and then is double-passed through an acousto-optic modulator (AOM) which is placed at the focuses of one pair of lens, acting as trapping laser. This design makes sure that the beam do not move spatially when its frequency is changed by its voltage controlled oscillator. The second laser frequency is locked to $F=3\rightarrow F'=3$ transition by SAS and is double-passed through another AOM. In our standard Cs MOT, the trapping laser and repumping laser frequencies are tuned to 15MHz below the $6S_{1/2}(F=4)\rightarrow 6P_{3/2}(F'=5)$ transition and $6S_{1/2}(F=3)\rightarrow 6P_{3/2}(F'=4)$ transition, respectively.

The available maximum trapping beam intensity is about 23mW/cm$^2$ and the available maximum repumping beam intensity is about 18mW/cm$^2$. The trapping and repumping beams are both expanded to about 15 mm in diameter. A charge-coupled device (CCD) is used to measure the cold atom dimension and fluorescence intensity which corresponds with cold atom number. Usually, the cold Cs atomic density and temperature in MOT are measured to be about $5\times10^{11}$/cm$^3$ and 70μK, respectively.

After accomplish of cold atoms in MOT, we will implement the PGC, release, and recapture processes to produce, release and detect the sub-Doppler cooling cold atoms, respectively. Fig.1 shows the timing sequence of total process. The cold atoms are sub-Doppler cooled in optical molasses by PGC after the precooling in MOT, and then expand freely in release process followed by recapture process, in which the cold atoms are recaptured by trapping and repumping beams. We do not show repumping laser on the figure as it has little influence on the cold atomic temperature in PGC. The PGC is a reliable and efficient method to achieve trapped cold atoms with sub-Doppler temperature. The most important step to perform PGC is that the cooling laser frequency must be shifted away from the near resonance transition to a larger detuning and cooling laser intensity must be turned down to a lower value. Usually, there are two frequently-used modes to change a physical value, which are step and ramp modes. So there are four combinations for cooling laser frequency and intensity, which are the 'frequency step+intensity ramp', 'frequency ramp+intensity step', 'frequency step+intensity step' and 'frequency ramp+intensity ramp'. We mark 'SR', 'RS', 'SS' and 'RR' for simplicity, which are shown in the dashed rectangle of Fig.1. In the release process, the trapping laser and repumping laser are both turned off in order to make the cold atoms expand freely. In the recapture process, the trapping and repumping lasers are suddenly turned on to recapture the cold atoms and CCD camera is triggered to record the fluorescence intensity and dimension which correspond the number and volume of cold atoms respectively. The chose gain and shutter time of CCD camera are 21dB and 1ms, respectively.

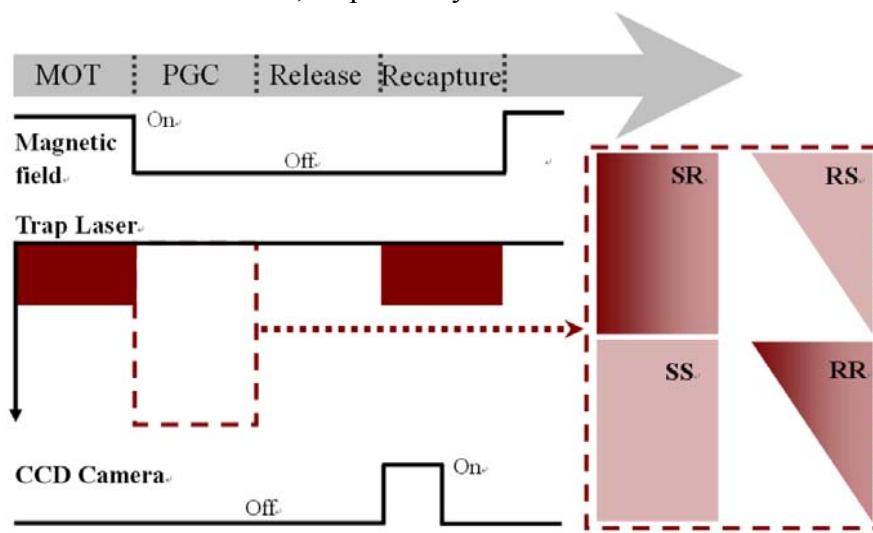

Fig.1 Timing sequence of sub-Doppler cooling of cold cesium atoms. The process contains magnetic-optical trap (MOT), polarization gradient cooling (PGC), release and recapture processes.

The process above is repeated at 10s period. A program based on Labwindows/CVI software is used to control the time sequence of magnetic field, trapping/cooling laser frequency and intensity, repumping laser intensity, and CCD acquisition. The AOM is used to change the trapping/cooling laser frequency and intensity through 'Freq in' and 'Mod in' modulator. The repumping laser intensity is also controlled by the 'Mod in' modulator of the other AOM. Usually, the current of coils which provide the magnetic field gradient can reach up to 50A in order to produce necessary gradient due to big dimension of stainless-steel chamber (MCF800-ExtOct-G2C8A16, Kimball Physics). An insulated gate bipolar transistor (IGBT; GA200SA60U) is used to switch the current on or off. The magnetic field can be completely turned on or off within 0.2ms according to our measurement.

## 3. Experimental results and discussions

### 3.1 The dependence of cold atomic temperature on the changing modes of cooling laser

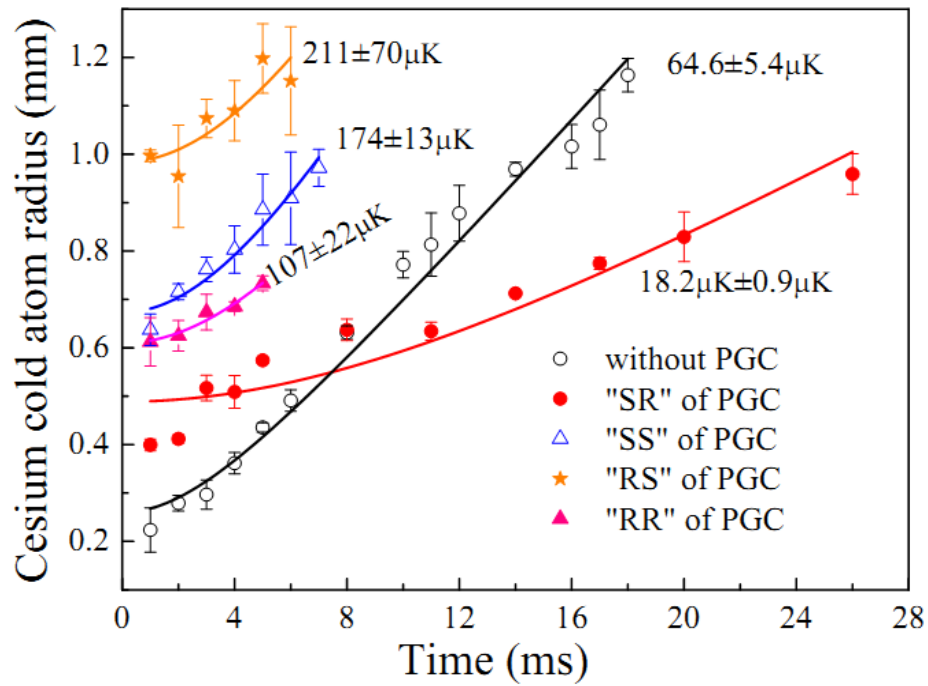

Fig.2 Expansion process of cold atoms in optical molasses without PGC and with PGC under four different changing modes of cooling laser. The labels 'S' and 'R' represent step and ramp modes, respectively. The first letter indicates the cooling laser frequency and the second letter indicates the cooling laser intensity. The solid lines are theoretical fitting curve through equation (2) in which the temperature can be obtained.

Just as described above, there are no literatures systematically investigating how the changing modes of cooling laser influence the cold atomic temperature. According to the description of experimental setup and timing sequence part, there are 'SR', 'SS', 'RS' and 'RR' modes to change the frequency and intensity of cooling laser. Fig.2 shows the expansion processes of cold atoms in optical molasses with PGC under the four different changing modes of cooling laser, comparing the one without PGC after MOT. The initial atom sample is the same for the five cases. The black open dots, red solid dots, blue open triangles, orange solid stars, and pink solid triangles represent the expansion processes without PGC, 'SR', 'SS', 'RS' and 'RR' modes with PGC, respectively. The error bars in the figure represent standard deviations in mean value of four averaged, repeated experimental data. The PGC interaction time is 6ms, the cooling laser intensity and frequency for each beam in PGC are 1mW/cm$^2$ and -10.5$\Gamma_{Cs}$ red detuning below the $6S_{1/2}(F=4) \rightarrow 6P_{3/2}(F'=5)$ transition, where $\Gamma_{Cs}$ is the natural linewidth. The repumping beam intensity is about 18mW/cm$^2$.

We use release and recapture method[18] to measure the cold atomic temperature after PGC process, which has been described in Fig.1. The temperature of cold atoms can be obtained from the expansion process by fitting this equation

$$\sigma^2(t) = \sigma^2(0) + \frac{k_B T}{M} t^2, \quad (2)$$

where $\sigma(t)$ and $\sigma(0)$ are the Gaussian radiuses of atomic cloud at $t$ and initial times respectively, $T$ and $M$ are the cold atomic temperature and atomic mass respectively. For simplicity we list the temperatures for different cases in Fig. 2. Comparing the cold atomic temperatures under 'SS' and 'RS' modes, which are 174μK and 211μK respectively, we found that step mode of cooling laser frequency is better than the ramp mode. Comparing the cold atomic temperatures under 'RR' and 'RS' modes, which are 107μK and 211μK respectively, we found that the ramp mode of cooling laser intensity is better than the step mode. It will be reasonable to predict that the 'SR' mode will be the best for PGC, that agrees with our measurement. Our measurement shows that the expansion process of cold atoms in 'SR' mode of PGC is the slowest, corresponding to the lowest temperature ~18μK under these experimental parameters.

We explain these experimental results as below. Usually, the intensities of six laser beams are not equal in cold atomic system. The reasons come from many aspects, such as the window gasses of chamber in our experiment will induce intensity loss for reflected cooling beams. The trapping laser frequency is detuned to about 3$\Gamma_{Cs}$ below the atom transition in MOT to cool atoms. When the cooling laser frequency is changed to several $\Gamma_{Cs}$ below atom transition by step mode, the cold atoms in PGC can be avoided to be heated by the cooling laser in PGC process. When the cooling laser intensity are changed to lower power by ramp mode, the atoms will be cooled adiabatically[19] as the atomic oscillation period in the dipole wells is small compared with the ramp time[20]. From the analysis, it is predicted that the 'SR' mode should be the best for PGC, which has been verified by our measurements. This result shows that 'SR' mode of cooling laser is the most efficient changing mode for polarization gradient cooling of cold atoms in optical molasses.

**3.2 The dependence of cold atomic temperature on PGC interaction time**

In the PGC process, the cold atom density will decrease as the interaction time increases. However too short time has little influence on sub-Doppler cooling of cold atoms. In order to find the appropriate interaction time, we investigate the dependence of the temperature of sub-Doppler cooling cold atoms on PGC interaction time. The experimental result is shown in Fig.3. The temperature is obtained by the mentioned release-recapture method. The error bars represent standard deviations in mean value of three averaged, repeated experimental data. The cooling laser changing mode is 'SR' and its frequency and intensity are still $-10.5\Gamma_{Cs}$ and $1mW/cm^2$. The repumping beam intensity is about $18mW/cm^2$.

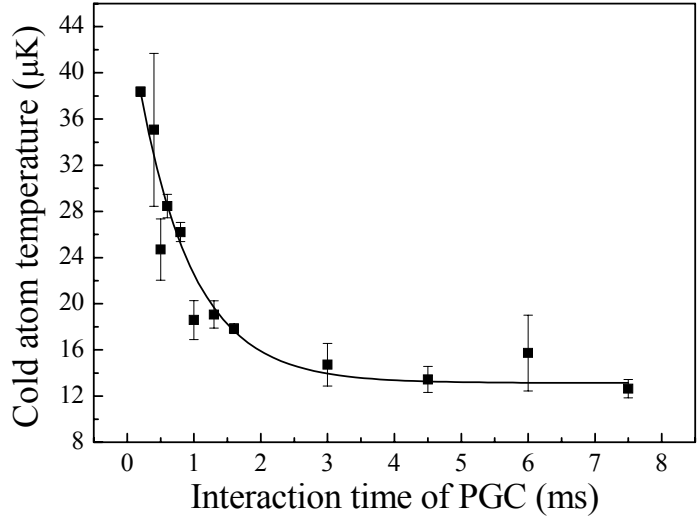

Fig.3 The dependence of cold atomic temperature on interaction time.

It shows that the temperature decreases rapidly from about 38μK to 15μK as time increases from 0.2ms to 3ms and reaches to a steady value after 3ms. We analyze the experimental result as below. It is reasonable to suppose that the temperature variation rate $\Delta T/\Delta t$ is proportional to the number of cold atoms $N$ which interact with the cooling laser. It needs to be declared that the number $N$ has only statistical significance. As the number $N$ is proportional to the temperature of cold atoms $T$, the temperature variation rate can be qualitatively described as:

$$\frac{dT}{dt} = -\frac{1}{\tau}T. \qquad (3)$$

Where $\tau$ is a characteristic time, the negative sign represents that the temperature decreases as time increases. We use the solution of this equation $T = T_0 \exp(-t/\tau)$ to fit the experimental data. The fitting result is shown with solid line in Fig.3. The characteristic time is 0.8ms, which means that the cycle times of atom and photon interaction are about several thousands in the characteristic time.

### 3.3 The dependence of cold atomic temperature on cooling laser frequency and intensity

As the equation (1) indicates that the temperature of cold atoms in PGC is influenced by cooling laser frequency and intensity, we will investigate the dependence of the sub-Doppler cooling cold atomic temperature on cooling laser frequency and intensity. The measured results are shown in Fig.4a and Fig.4b. The temperature is also obtained by the release-recapture method. The error bars represent standard deviations in mean value of three averaged, repeated experimental data. The cooling laser changing mode is 'SR' and the PGC interaction time is 1.5ms. The repumping beam intensity is about 18mW/cm$^2$. The cooling laser intensity and frequency for each beam are about 1mW/cm$^2$ and -8.6$\Gamma_{Cs}$ respectively for Fig.4a and Fig.4b.

Fig.4a shows that the temperature decreases rapidly from about 150μK to about 20μK as the red detuning of cooling beam changes from -4$\Gamma_{Cs}$ to -6$\Gamma_{Cs}$. This is the expected results from Sisyphus cooling[8] which is the cooling mechanism of PGC. The temperature then reaches to a steady value when the cooling laser frequency is out of -6$\Gamma_{Cs}$ red detuning.

Fig.4b shows that the temperature increases proportionally from about 19μK to 24μK as the cooling laser intensity for each beam increases from about 1mW/cm$^2$ to 6.7mW/cm$^2$. This is consistent with the equation (1). However, the temperature becomes large when the cooling laser intensity is lower than 1mW/cm$^2$. This is due to the heating effect from cooling laser. As we have announced, the intensities of six laser beams are not equal in our system. The atoms in PGC are heated when the cooling laser intensity is lower than the saturation intensity of Cs atom for σ$^\pm$ polarized light, which is 1.1mW/cm$^2$[21].

In order to verify our explanation, we measure the temperature of cold atoms in PGC as a function of repumping laser intensity in PGC, which is shown in Fig.5. The cooling laser changing mode is 'SR' and the interaction time is 1.5ms. The cooling laser frequency and intensity are -10.5$\Gamma_{Cs}$ and 1mW/cm$^2$. It shows that the temperature does not change at large reumping laser intensity. However, the temperature increases as the intensity is smaller than about 6mW/cm$^2$, which is thought to arise from the heating effect of repumping laser. The repumping laser has no contribution to PGC process and has only one beam, thence the heating effect is obvious at a larger intensity, comparing the intensity of cooling laser. When the intensity is around the saturation intensity(1.1mW/cm$^2$), the temperature reaches a relatively larger value.

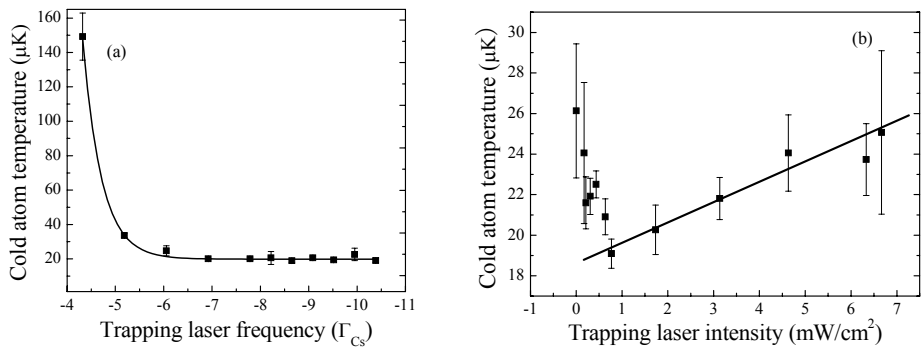

Fig.4 The dependence of cold atomic temperature on the cooling laser frequency and intensity.

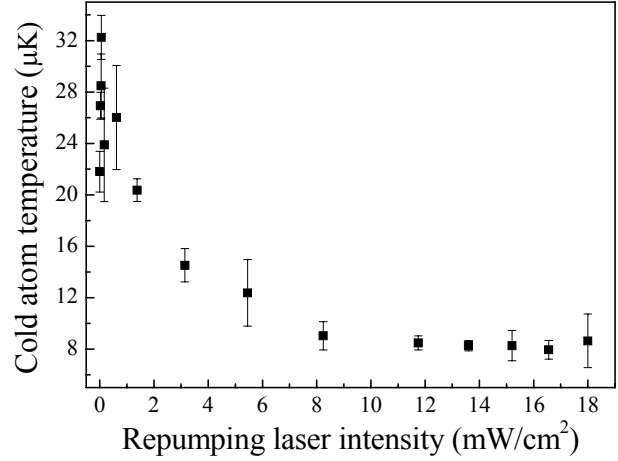

Fig.5 The dependence of cold atomic temperature on repumping laser intensity.

In cold atomic experiments, we usually take attentions on not only the temperature, but also the number, density of cold atoms, and phase space density. Finally, the optimal parameters for polarization gradient cooling of $^{133}$Cs are chose as this: changing mode of cooling laser is SR, PGC interaction time is 2ms, cooling laser frequency and intensity in PGC is $-6\Gamma_{Cs}$ and 1mW/cm$^2$ respectively. Using these parameters, we capture a sample of cold atoms with the temperature of 4.2±0.6μK, the number of $2\times10^9$, the density of $1\times10^{11}$/cm$^3$, and phase space density of $4.4\times10^{-5}$.

## 4. Conclusion

We have systematically investigated the dependence of the temperature of cold cesium atoms in optical molasses by polarization gradient cooling on controllable experimental parameters, which contain changing modes of cooling laser, PGC interaction time, cooling laser frequency and its intensity. Our results showed that the 'SR' mode of cooling laser, which means cooling laser frequency is stepped to the final value while the cooling laser intensity is ramped to the final value, is the best for the polarization gradient cooling of cold atoms. The reason is that cold atoms can avoid the heat effect from cooling laser and can be cooled adiabatically by this mode. It is shown that the cold atomic temperature in PGC decreases rapidly in the first third milliseconds and reaches to a steady value after 3ms. We introduce a statistical explanation and use an exponential decay formula to fit the experimental data. The variation of cold atomic temperature in PGC on the cooling laser frequency is similar as the one on the interaction time. It is predicted that the temperature increases proportionally with trapping laser intensity in the range of 1-7mW/cm$^2$. However, it is somewhat unusual that the temperature becomes large when the cooling laser intensity is lower than 1mW/cm$^2$, which is around the saturation intensity of Cs atom for $\sigma^{\pm}$ polarized light. It is due to the heating effect from cooling laser with lower intensity than the saturation intensity. The explanation has been verified by the measurement of temperature on the repumping laser intensity. After optimization, the lowest temperature of cold cesium atoms is observed to be about 4μK with the number of $2\times10^9$, a density of $1\times10^{11}/cm^3$, and a phase space density of $4.4\times10^{-5}$. The optimization process can help us to more deeply understand the interaction mechanism of cooling laser and cold atoms in PGC and experimental parameters are also meaningful for other cold atoms.

## Acknowledgments

This work was supported by the National Basic Research Program of China (973 Program, Grant No. 2012CB921603), the International Science & Technology Cooperation Program of China (Grant No. 2011DFA12490), the National Natural Science Foundation of China (Grants No. 61275209, 10934004 and 11004125) and NSFC Project for Excellent Research Team (Grant No. 61121064).